# Knowing When to Splurge: Precautionary Saving and Chinese-Canadians

*Abstract:* Why do household saving rates differ so much across countries? This micro-level question has global implications: countries that systematically "oversave" export capital by running current account surpluses. In the recipient countries, interest rates are thus too low and financial stability is put at risk. Existing theories argue that saving is precautionary, but tests are limited to cross-country comparisons and are not always supportive. We report the findings of an original survey experiment. Using a simulated financial saving task implemented online, we compare the saving preferences of a large and diverse sample of Chinese-Canadians with other Canadians. This comparison is instructive given that Chinese-Canadians migrated from, or descend from those who migrated from, a high-saving environment to a low-savings, high-debt environment. We also compare behavior in the presence and absence of a simulated "welfare state", which we represent in the form of mandatory insurance. Our respondents exhibit behavior in the saving task that corresponds to standard economic assumptions about lifecycle savings and risk aversion. We find strong evidence that precautionary saving is reduced when a mandatory insurance is present, but no sign that Chinese cultural influences – represented in linguistic or ethnic terms – have any effect on saving behavior.

PsycINFO classification 3920    JEL classification D14, D15, G14

**Corresponding author:**

Mark S. Manger  
Associate Professor of Political Economy  
Munk School of Global Affairs & Public Policy  
University of Toronto  
Toronto, ON  
mark.manger@utoronto.ca

J. Scott Matthews  
Associate Professor  
Department of Political Science  
Memorial University of Newfoundland,  
St. Johns, NL  
scott.matthews@mun.ca

*Acknowledgements*

The authors would like to thank Thomas Sattler, Evelyn Hübscher, Michael Cemerin, Jonas Markgraf, Gabriele Spilker, Mark Pickup and Eric Wang for helpful advice and comments, and Tanvi Shetty for excellent research assistance. All remaining errors are ours.

*Funding*

This research was funded by SSHRC Grant 435-2014-0487.

# Introduction

饭疏食饮水，曲肱而枕之，乐亦在其中矣。

The Master said, To eat coarse greens, drink water,

and crook one's elbow for a pillow – joy also lies therein.

*Confucius, The Analects, Book VII.*

Why do people in some countries save large shares of their disposable income, while people in other countries choose to incur considerable debt through mortgages, car loans and credit cards? Household saving rates differ widely across countries, ranging from 20 percent of GDP in China and 15 percent in Hong Kong to less than 5 percent in the United States and below zero in New Zealand. This micro-level issue has large – indeed, global – implications. Unless all of these savings are used for investment purposes, a high-saving country has a surplus of capital. Countries that systematically save more than they spend export this capital by running current account surpluses. In the recipient countries, interest rates are therefore lowered and households are likely to take on even more debt. Savings and debt are therefore among the primary determinants of the stability of financial systems. Growing capital exports have been called a "savings glut" that has fueled bubbles and price distortions (Bernanke 2005), and carry much of the blame for the recent global financial crisis.

Yet despite the importance of the determinants of household saving in macroeconomics and policy-making, our understanding of cross-country differences is extremely limited. While economists regularly assume that savings is precautionary and serves to protect against income shocks, this assumption has not been substantiated with experimental data. It is therefore an open question if the differences in savings are a largely rational response to institutional constraints, or rather due to cultural practices and social learning.

In this paper, we present the findings from an online original survey experiment using a large (n = 1,214) sample of the Canadian population. We ask two questions. First, does mandatory insurance reduce savings? In other words, to what extent is saving *precautionary*, as generally assumed in the economic literature? Second, we investigate the saving behavior of an ethnocultural subgroup in our sample: do people of Chinese descent save more than other population groups, as an important strand of literature and observational data suggest? In other words, are there *cultural* driving forces that would make Chinese "excess saving" resistant to policy changes? By design, our sample contains a considerable oversample of Canadians of Chinese descent and, thus, we are in a unique position to detect cultural differences as may exist. As a direct implication we also study whether mandatory insurance lowers savings differently for Chinese immigrants and their descendants compared to other groups in our sample.[1]

We find strong evidence that precautionary saving is reduced when a mandatory insurance is introduced. This treatment effect is fairly large and approximates 20 percent of average savings. Our results also speak against any predisposition of people of Chinese descent to save more than other ethnic groups. We find no evidence that respondents of Chinese descent in our sample save more than others, neither unconditionally nor when controlling for a variety of demographic and economic factors. These findings are robust to different definitions of Chinese descent; in particular, they hold when we define "Chinese" in both linguistic and ethnic terms. We also find very limited evidence of Chinese cultural differences between treatment subgroups, after controlling for demographic and economic factors. Precautionary saving does not seem to be greatly influenced by cultural or linguistic factors in our experiment.

Our study makes several contributions. First, our results count among a small number of experimental investigations of the relationship between precautionary saving and mandatory insurance, a topic that has thus far mostly been explored with observational data. Our findings suggest that the widespread assumption that much saving is precautionary is strongly justified empirically. Second, we demonstrate that to the extent that cultural or sociological elements affect saving behavior, they do not make people of Chinese descent more likely to save in our experiment. Third, given that the treatment effect of mandatory insurance is large and precisely estimated, our findings have policy implications for the "rebalancing" of the Chinese and other surplus economies toward greater consumption: if there is little evidence of cultural effects, then a strengthening of social safety nets would be an effective policy choice to reduce precautionary saving levels.

---

[1] Our choice of self-identified people of Chinese descent is primarily driven by policy relevance—the observation of extremely high household saving rates in all Chinese-language jurisdictions—and convenience, as their relatively large presence in Canada allows us to obtain a sufficient sample. Of course, other groups that exhibit higher than average saving (e.g. Germans) could equally stand in for a test of cultural factors shaping behavior.

## Precautionary Saving in China and Elsewhere

The striking variability of household saving rates across countries in general and the remarkably high saving rates in some East Asian countries in particular have stimulated a rich debate. Even in the East Asian context, China stands out with gross saving reaching 50 percent of GDP between 2007 and 2009 and household saving contributing an historically unprecedented 20 percent of GDP (Ma and Wang 2010). The literature falls into two camps. The first grants no role to culture or socialization and assumes that individual-level Chinese saving behavior is driven by the same factors as saving elsewhere: a bequest motive and precautionary saving. The second gives varying degrees of weight to variables of "Chinese-ness", such as habit, culture and even language.

Among the first set of scholars, many authors attribute the increase in Chinese saving rates over the last decades to rising income uncertainty (Chamon, Liu, and Prasad 2013) and an increasing burden of housing, education and health care (Chamon and Prasad 2010). Chan and collaborators (2016) suggest that increasing economic inequality leads to underconsumption if capital markets are underdeveloped, as in China, because the poor cannot borrow from the rich. Choi, Lugauer and Mark (2014) estimate household preferences and find that 80 percent of Chinese saving is motivated by precautionary motives; the comparable figure for the United States is roughly half. Yao et al. (2011) use survey reports to document that Chinese households (in China) are more likely to save for precautionary and education purposes than American households, while saving has clear lifecycle patterns. This fits well with another study finding that the introduction of National Health Insurance (NHI) reduced saving levels in Taiwan (Kuan and Chen 2013).

Many authors submit that most of the increase in Chinese saving is in fact due to demographic shifts and economic growth (Bonham and Wiemer 2013; Chao 2011; Choukhmane, Coeurdacier, and Jin 2013; Curtis, Lugauer, and Mark 2011, 2011; Kraay 2000;

Yang, Zhang, and Zhou 2011). Interestingly, this discussion echoes earlier research on Japan (See Horioka 2010 for a review) and fits well with cross-national studies (Horioka and Terada-Hagiwara 2012; Loayza, Schmidt-Hebbel, and Servén 2000).

Yet these arguments are not without detractors. Conqvist and Siegel (2015) use twin studies to argue that up to a third of the variation in saving behavior is genetic in its origins. Horioka and Wan (2007) suggest that "inertia" (i.e. past saving rates) explains most of the saving in China. Wei and Zhang (2011) submit that unfavorable sex ratios force Chinese young men (or their parents) to save more to become credible marriage prospects. In a much-cited cited paper, Chen (2013) even provides econometric evidence that having a native language without a "strong future tense" (like Chinese and its various dialects) is strongly correlated with higher household saving rates across the world, findings echoed by Guin (2016), who compares French and German language households in Switzerland. A significant number of scholars therefore reserve a role for cultural factors, broadly defined, that would predispose some people to a greater propensity to save.

More generally, claims that differences in saving behavior are due to socialization and inter-generational transmission are common (Almenberg et al. 2018), and the low saving rates of American households are often attributed to a "consumerist culture" (Garon 2011). In other research, materialist values are correlated with a propensity to incur debt rather than save (Watson 2003).

While insightful, these works face a fundamental challenge: even country-level surveys necessarily hold all institutional and macro-cultural variables, however defined, constant. The alternative is, thus, to run experiments with individuals. There are relatively few such studies, and they tend to be focused on testing whether individual behavior fits the assumptions of economic optimization models (Carbone and Duffy 2014; Meissner 2014). One study in particular uncovers that precautionary saving may be learned (Ballinger, Palumbo, and Wilcox 2003), in the sense that when it is observed, subjects learn to perform it more consistently. Others find that positional goods ("keeping up with the Joneses") negatively affect savings (Feltovich and Ejebu 2014). These studies generally indicate considerable individual-level variation in the ability to optimize.

## Investigating Precautionary Saving Experimentally

We investigate three conjectures in our experiment. First, and most prominently, we examine whether a mandatory insurance treatment reduces saving. In our simulated financial saving task, described in more detail below, our respondents encounter the possibility of expense shocks while trying to smooth consumption. This forces them to accumulate savings as a cushion against large expenses. If saving is precautionary, we therefore expect the mandatory insurance treatment to reduce saving.

Second, we ask whether Chinese-descent Canadians, including those who immigrated and those who are at least first-generation Canadian (i.e. born in Canada), differ in their saving behaviour from other Canadians. As noted above, we employ two definitions of "Chinese descent" – we ask respondents to self-identify as having at least one Chinese grandparent, or growing up in a Chinese-speaking household, where "Chinese language" is defined as Standard Chinese (Mandarin), Cantonese, or another Chinese dialect such as Hokkien. Nearly all of our respondents in this category in fact have four Chinese grandparents and two Chinese parents.

Finally, we study if there is a difference in treatment effects across groups, based on the conjecture that a learned or cultural propensity to save – as a, presumably, stable psychological predisposition – should be less affected by a mandatory insurance treatment.

*H1. The mandatory insurance treatment should reduce savings (precautionary motive).*

*H2. Ethnic Chinese will save more than other groups (culture).*

*H3. The negative effect of the treatment condition on savings will be smaller for ethnic Chinese than others (moderated precautionary motive).*

## Experimental Design

Our experimental approach employs a particular version of the lifecycle consumption model. There is a finite world with a discrete number of periods *T*. The decision-maker receives an income in each period that rises over time, experiences involuntary and variable expenses, and must decide how much to consume and save. Saving does not earn any interest, and savings left over at the horizon become worthless, which excludes the bequest motive. Following a phase of income earning is a time period, akin to "retirement", in which consumption is based on drawing down savings, but during which subjects can still experience expense shocks. Subjects can therefore work out by backward induction what the optimal consumption path would be—although, given the computational challenges, we except them to do so only approximately (Ballinger et al. 2011).

We furthermore induce consumption-smoothing by offering a reward for subjects to incentivize them to consciously avoid fluctuations in spending. Unbeknownst to subjects, they all receive the same income and expenses over a lifecycle, but in a randomly selected order. We can therefore expect that as the sample size grows, the temporal order of income shocks ceases to matter for our estimates. All subjects face the same incentives on average, allowing us to identify the treatment effect.

Our dependent variable is the savings left over at the end of retirement, i.e. the share of income received during income-earning periods that is put away for later periods after expenses and consumption, observed at the start of the last period of the life cycle (i.e., period 24; see below).

### Implementation

Our simulation builds on the saving task proposed by Koehler and collaborators (2015), but introduces an experimental manipulation. In the simulation, respondents are given the task to allocate spending and saving during four separate and independent "lifecycles" consisting of 24 periods each. Respondents receive an income stream – which starts at 10 units per period and grows by 1 unit every two periods – for 18 periods, which is followed by 6 periods of "retirement" during which they must draw down their savings. This allocation is known and shown on the screen.

In each period, participants can consume any amount that does not exceed their current monetary units, which consist of savings plus income from that period. Participants are instructed that the lower the variability of their spending over the lifecycle (i.e., the smoother their consumption) the better their chances of winning a prize (a $50 Starbucks gift card). Participants cannot deliberately choose to go into debt, but if the sum of savings and current

period income is less than the expenses drawn in the period, participants go into involuntary debt, and all income in the next period is automatically applied to the debt first before units are available for consumption. This implies that if a participant has zero savings and ten units of income, and an expense of ten units is drawn, they will have zero units available in the subsequent period. It also implies that to end the game without debt, they must make some precautionary savings. Any unspent units, finally, are automatically spent in the last period.

Respondents' understanding of the saving task is tested with a set of multiple-choice questions. We filter respondents so that only those who answer the multiple-choice questions correctly and have understood the mechanisms of the game can proceed, although respondents have as many tries as they like to do so. To allow for some margin of error in the decision-making, the game is played for four "lifecycles" and the values are averaged over the four periods.

In each period, participants face a randomly drawn level of expenses. They are shown a distribution of possible expenses (ranging from 2 to 10) and shown that "2" is "very, very likely" and "10" "very, very unlikely" to occur in each period. Unbeknownst to participants, the expenses are drawn without replacement—e.g. once an expense of 10 has occurred it cannot be incurred a second time. Possible expenses are once 10, 9, 8, 7, 6 each, 3 x 5, 3 x 4, 4 x 3, 5 x 2. Accordingly, all subjects receive the same income over time and an identical total of expense shocks, but the order of these shocks varies randomly.

The insurance treatment implies that respondents are informed that, in each period, a mandatory "insurance premium" of 2 units will be deducted from their income. If they experience an expense shock between 7 and 10 units, the insurance covers 5 of these units. Expense shocks of 5 or less have to be covered out of savings. The co-payment in both scenarios creates "residual uncertainty," without which it would be irrational to adjust saving patterns.

To make the game both comprehensible and the numbers manageable in mental calculations, we use only small round numbers. Our dependent variable, as noted, is the final sum of savings at the end of a lifecycle. An inevitable consequence of using round numbers is that disposable income is about 7 percent smaller in the treatment group. We correct this by adjusting upwards the savings level in the treatment group. Specifically, in the treatment condition, savings observed in period 24 are divided by the ratio of expected income in the treatment to expected income in the control group (approximately .93).

The appendix shows the saving simulation as presented to respondents on their computer screen. We preface the saving task itself with a survey that collects demographic information. Prior to completing the survey, respondents can choose if they want to complete the study in English, Chinese (simplified characters) or Chinese (traditional characters). A relatively small number of respondents chose one of the two varieties of Chinese.

### Characteristics of the Sample

Our sample consists of 1,214 adult Canadians living in 11 provinces or territories, with close to half of our respondents based in Ontario and a quarter from British Columbia. Quebec residents are underrepresented as we fielded our survey instrument only in English and Chinese. We ask respondents to self-identify as immigrants of Chinese descent (born in China, Taiwan or Hong Kong), Canadian-born of Chinese descent (at least one grandparent born in China, Taiwan or Hong Kong), immigrants not of Chinese descent (no grandparent born in China, Taiwan or Hong Kong), or Canadian-born not of Chinese descent (no grandparent born in China, Taiwan or Hong Kong). We oversample those of Chinese descent such that fully 68.4 percent of our respondents fit that description, and half of these respondents (50.3 percent) are immigrants to Canada. We also gather information about various demographic characteristics. The typical respondent in our sample (i.e., who takes on the sample mode of each of the noted characteristics) is a woman (49.5 percent) who lives in Ontario (48.9 percent) and completed a Bachelor's degree (50.5 percent), has a household income between 50 and 100 thousand dollars per year (36.3 percent), has net household wealth of greater than $200,000 (48.8 percent), and is between 26 and 44 years old (46.9 percent[2]). Note that the distributions of these variables differ somewhat between those of Chinese descent and others (all Chi-square values exceed the 90-percent confidence level or better); accordingly, we control for these characteristics in estimating cultural differences in saving behavior, below.

---

[2] Age, which we measure in 7 categories, is effectively bimodal, with 23.7 percent aged 26 to 34 years and 23.2 percent aged 25 to 44 years.

## Analysis

Given the design of our study, each participant yields four separate observations: one for each completed "lifecycle" of the saving task. Accordingly, we stack the data such that each respondent is represented by four rows in the dataset, one for each cycle. We estimate OLS regressions including fixed effects for cycles, and with standard errors corrected for clustering within respondents. Our dependent variable ranges from -57 to 140 with a mean of 27.2, and is approximately normally distributed.

Given the considerable individual-level, systematic variation in saving behavior, we control for a variety of demographic variables suggested by previous research (n.b., levels of categorical variables indicated in Table 1): gender (woman=1), age (7 categories), province (Ontario v. British Columbia v. all others), education (4 categories), income (6 categories), and wealth (6 categories). Note that, as a sizable share of respondents declined to report their level of income [~14 percent] or wealth [~10 percent], we separately identify these respondents in all regressions. Finally, given that individual risk attitudes are also likely determinants of saving behavior (Carbone and Duffy 2014), we also include in all models the financial risk aversion subscale of the Domain-Specific Risk-Taking (DOSPERT) scale for adult populations. There is some suggestive evidence that Chinese with higher incomes (>$50k/year) are more willing to take financial risks, as shown in Figure 1.

Table 1 shows the full regression results, while Figure 2 plots the coefficient estimates of theoretical interest in each model. Overall, we find strong support for the precautionary motive hypothesis (H1), but only very limited evidence of cultural effects, whether direct in nature (H2) or in interaction with the insurance treatment (H3).

Model (1) is our basic model, including controls for the DOSPERT-F subscale measure, gender, age, province of residence, education, income level and personal wealth. Inspecting the estimates for these controls serves as a check on the validity of the assumption that behavior observed in the saving game is a suitable analog to real-world saving behavior.

Importantly, the DOSPERT-F scale functions as expected, with risk-takers saving considerably less than the risk-averse. More specifically, the estimates imply that those at the top of the scale save 13.5 fewer units, by the final period, than those at the bottom of the scale.[3] This effect is fully one-half the sample average of savings in the final period (i.e., 13.542/27.248 = .505). The effects of gender and age are also highly sensible. Women save in excess of 8 units more than men, on average. There is a lifecycle in saving (see Figure 2) such that saving increases until middle age (more precisely, the 45-54 yrs. age category) then begins to decline.[4] Less readily interpretable are the negative effects of Ontario residence and the holding of (only) a Bachelor's degree, and the positive effect on saving of refusing to report one's household income. We also note that, within respondents, saving levels increase after the first cycle and are fairly stable thereafter, which may reflect a learning process (i.e., some respondents in the first cycle may retain insufficient units to cover expenses in the no-income periods, and thus adjust their behavior in subsequent cycles).

Of primary interest, however, is the estimated effect of the insurance treatment. In this specification, the difference in (adjusted) final period savings is nearly 6 units, an effect that is statistically significant at the 99.9-percent level. The estimate implies a substantial negative effect of the treatment on saving, one that is more than one-fifth the size of the sample average level of savings (i.e., 5.968/27.248 = .219) and nearly half the size of the effect of a shift across the range of the risk-taking scale (i.e., 5.968/13.542 = .441). The result implies that precautionary motives may be an important source of saving behavior.

Model (2) adds the measure of Chinese ethnicity. Contrary to expectations, we detect no influence of this variable on saving behavior: the coefficient estimate is tiny and its standard error is more than seven times larger. Model (4) tests the cultural hypothesis using Chinese speaking (defined as "growing up in a household where the primary language was Chinese"), rather than ethnicity, as the indicator – again, we detect no significant effect.

Models (3) and (5) test the conditionality of the insurance treatment's effect on cultural variables, using the ethnicity and language indicators, respectively. While there is no evidence of an interaction between Chinese ethnicity and the insurance condition, the interaction between Chinese speaking and the insurance condition is sizable and in the expected direction. In a one-tailed test, furthermore, the interaction is significant at the 95-percent level ($p = .049$).

---

[3] The DOSPERT-F scale ranges from 1 to 7. Thus, this effect is six times the coefficient estimate.
[4] There is a hint that saving may increase after age 75, but the estimate is far from conventional significance thresholds ($p = .220$) and reflects the behaviour of just 10 respondents.

The interaction implies that the negative effect of the insurance condition is considerably larger among non-Chinese speakers: the "main effect" of the treatment implies a reduction of 8.3 units in saving in this group. Among Chinese speakers, however, the treatment effect is nearly two-thirds smaller. It is possible, for example, that growing up in a Chinese-speaking household – rather than simply having Chinese origins, but growing up in an English-speaking cultural milieu – is required to transmit a cultural preference for thrift. While this result is consistent with expectations, the estimate's imprecision, and the absence of a similar effect of Chinese ethnicity, renders the evidence for H3 suggestive at best.

## Conclusion

Saving rates vary widely across countries, but given the multitude of factors that influence financial decisions, it is difficult to study saving behavior in isolation of cultural values and formal and informal social and economic institutions. Most research therefore simply assumes that a lot of saving is precautionary. In this paper we offer experimental evidence that this appears to be a valid assumption. In our simulated financial saving task, subjects behave very close to what these assumptions would lead us to expect, with a clear lifecycle effect, and a sizable and statistically significant reduction in saving levels when subjects are treated with mandatory insurance against major expenses.

    Controlling for various demographic and economic factors yields sensible coefficients, so that we have some confidence that our results are not just characteristics of our sample. Meanwhile, we do not find any robust evidence that one particular ethnic group, people of Chinese descent in Canada—whose ethnic confrères in China, Hong Kong and Taiwan all exhibit extremely high personal saving rates—is more disposed to saving than other subjects in our sample. These findings come with obvious caveats. Chinese-Canadians either selected into, or descend from those who selected into, immigration, although not always voluntarily, and Chinese of different generations may have emigrated for very different reasons. Nonetheless, if there was a pronounced Chinese cultural characteristic of frugality for precautionary reasons, it is improbable that we find no effect of this in our experiment.

Our findings, to the extent that they have external validity, have important policy implications. Much of the debate about Chinese current account surpluses and bilateral imbalances with the United States can be traced back to underconsumption by the Chinese. In macroeconomic terms, this is a necessity to achieve an external surplus. Our findings suggest that this behavior is driven by economic incentives and institutions, and that given the limited effect of cultural practices (if there is any at all), changing these incentives would likely reduce the very high savings rates. As much saving is precautionary in motive, according to our findings, an expansion of various public, mandatory insurance mechanisms would likely reduce excess household savings and contribute to a successful rebalancing of the Chinese economy, as well as those of other high-saving Chinese jurisdictions.

Ma, Guonan, and Yi Wang. 2010. "China's High Saving Rate: Myth and Reality." *Economie Internationale* 122(2): 5–39.

Meissner, Thomas. 2014. "Intertemporal Consumption and Debt Aversion: An Experimental Study." http://www.econstor.eu/handle/10419/100522 (February 27, 2015).

Watson, John J. 2003. "The Relationship of Materialism to Spending Tendencies, Saving, and Debt." *Journal of Economic Psychology* 24(6): 723–39.

Wei, Shang-Jin, and Jiaobo Zhang. 2011. "The Competitive Saving Motive: Evidence from Rising Sex Ratios and Savings Rates in China." *Journal of Political Economy* 119(3): 511–64.

Yang, Dennis Tao, Junsen Zhang, and Shaojie Zhou. 2011. *Why Are Saving Rates so High in China?* Cambridge, MA: National Bureau for Economic Research. http://www.nber.org/papers/w16771.

Yao, Rui, Feifei Wang, Robert O. Weagley, and Li Liao. 2011. "Household Saving Motives: Comparing American and Chinese Consumers." *Family and Consumer Sciences Research Journal* 40(1): 28–44.
15

Table 1. Experimental, demographic and risk preference effects on adjusted final period savings

|  | (1) | (2) | (3) | (4) | (5) |
|---|---|---|---|---|---|
| **Insurance condition** | -5.968 | -5.974 | -6.233 | -5.966 | -8.336 |
|  | (0.000) | (0.000) | (0.031) | (0.000) | (0.000) |
| **Chinese ethnicity** |  | -0.250 | -0.389 |  |  |
|  |  | (0.891) | (0.868) |  |  |
| **Chinese speaker** |  |  |  | -0.231 | -2.033 |
|  |  |  |  | (0.886) | (0.336) |
| **Ethnicity × Insurance** |  |  | 0.382 |  |  |
|  |  |  | (0.911) |  |  |
| **Lang. × Insurance** |  |  |  |  | 5.142 |
|  |  |  |  |  | (0.098) |
| **DOSPERT-F** | -2.257 | -2.241 | -2.242 | -2.251 | -2.230 |
|  | (0.003) | (0.003) | (0.003) | (0.003) | (0.003) |
| **Cycle** |  |  |  |  |  |
| 2 | 3.608 | 3.600 | 3.600 | 3.608 | 3.605 |
|  | (0.000) | (0.000) | (0.000) | (0.000) | (0.000) |
| 3 | 4.097 | 4.078 | 4.078 | 4.097 | 4.094 |
|  | (0.000) | (0.000) | (0.000) | (0.000) | (0.000) |
| 4 | 3.842 | 3.843 | 3.843 | 3.843 | 3.839 |
|  | (0.000) | (0.000) | (0.000) | (0.000) | (0.000) |
| **Woman** | 8.195 | 8.204 | 8.204 | 8.194 | 8.210 |
|  | (0.000) | (0.000) | (0.000) | (0.000) | (0.000) |
| **Age** |  |  |  |  |  |
| **26-34** | 6.699 | 6.754 | 6.762 | 6.686 | 6.971 |
|  | (0.020) | (0.021) | (0.021) | (0.020) | (0.016) |
| **35-44** | 7.510 | 7.529 | 7.536 | 7.499 | 7.633 |
|  | (0.008) | (0.009) | (0.009) | (0.008) | (0.007) |
| **45-54** | 10.215 | 10.231 | 10.236 | 10.176 | 10.462 |
|  | (0.001) | (0.001) | (0.001) | (0.001) | (0.001) |

Table 1 (cont'd)

| | | | | | |
|---|---|---|---|---|---|
| **55-64** | 7.331 | 7.327 | 7.332 | 7.301 | 7.551 |
| | (0.016) | (0.019) | (0.019) | (0.017) | (0.013) |
| **65-74** | 7.516 | 7.485 | 7.488 | 7.463 | 7.692 |
| | (0.111) | (0.118) | (0.118) | (0.113) | (0.102) |
| **75 or older** | 13.599 | 13.667 | 13.660 | 13.569 | 13.063 |
| | (0.220) | (0.219) | (0.219) | (0.223) | (0.247) |
| **Ontario** | -4.070 | -4.060 | -4.057 | -4.049 | -4.085 |
| | (0.042) | (0.044) | (0.044) | (0.044) | (0.042) |
| **BC** | -1.559 | -1.576 | -1.573 | -1.528 | -1.520 |
| | (0.498) | (0.500) | (0.501) | (0.510) | (0.511) |
| **Education** | | | | | |
| **Bachelor's** | -5.150 | -5.114 | -5.111 | -5.108 | -5.111 |
| | (0.010) | (0.013) | (0.013) | (0.012) | (0.012) |
| **Graduate/Prof'l** | -3.187 | -3.142 | -3.136 | -3.156 | -3.239 |
| | (0.204) | (0.216) | (0.217) | (0.209) | (0.197) |
| **Ph.D.** | -5.783 | -5.756 | -5.750 | -5.759 | -5.940 |
| | (0.154) | (0.158) | (0.159) | (0.156) | (0.144) |
| **Income** | | | | | |
| **25-50 k** | 3.288 | 3.494 | 3.477 | 3.304 | 3.122 |
| | (0.397) | (0.373) | (0.378) | (0.395) | (0.422) |
| **50-100 k** | 1.436 | 1.607 | 1.586 | 1.435 | 1.195 |
| | (0.687) | (0.655) | (0.662) | (0.687) | (0.738) |
| **100-200 k** | 1.005 | 1.280 | 1.265 | 1.014 | 0.847 |
| | (0.792) | (0.739) | (0.743) | (0.789) | (0.824) |
| **>200 k** | 3.378 | 3.602 | 3.601 | 3.371 | 3.348 |
| | (0.507) | (0.481) | (0.481) | (0.508) | (0.509) |
| **Refused** | 11.621 | 11.968 | 11.939 | 11.597 | 11.235 |
| | (0.009) | (0.008) | (0.009) | (0.010) | (0.012) |
| **Wealth** | | | | | |
| **25-50 k** | 5.335 | 5.291 | 5.298 | 5.354 | 5.564 |
| | (0.250) | (0.255) | (0.255) | (0.248) | (0.231) |

Table 1 (cont'd)

| | | | | | |
|---|---|---|---|---|---|
| **50-100 k** | -2.605 | -2.627 | -2.631 | -2.566 | -2.586 |
| | (0.480) | (0.480) | (0.479) | (0.487) | (0.484) |
| **100-200 k** | 1.816 | 1.780 | 1.787 | 1.841 | 1.911 |
| | (0.607) | (0.619) | (0.619) | (0.603) | (0.590) |
| **>200 k** | 1.865 | 1.853 | 1.853 | 1.923 | 1.862 |
| | (0.554) | (0.562) | (0.562) | (0.546) | (0.559) |
| **Refused** | -2.539 | -2.705 | -2.704 | -2.475 | -2.416 |
| | (0.473) | (0.453) | (0.453) | (0.487) | (0.499) |
| **Constant** | 24.443 | 24.347 | 24.454 | 24.462 | 25.261 |
| | (0.000) | (0.000) | (0.000) | (0.000) | (0.000) |
| **R-squared** | 0.060 | 0.060 | 0.060 | 0.060 | 0.062 |
| **N** | 4820 | 4800 | 4800 | 4820 | 4820 |

Note: OLS regression estimates. Standard errors corrected for clustering within individuals.

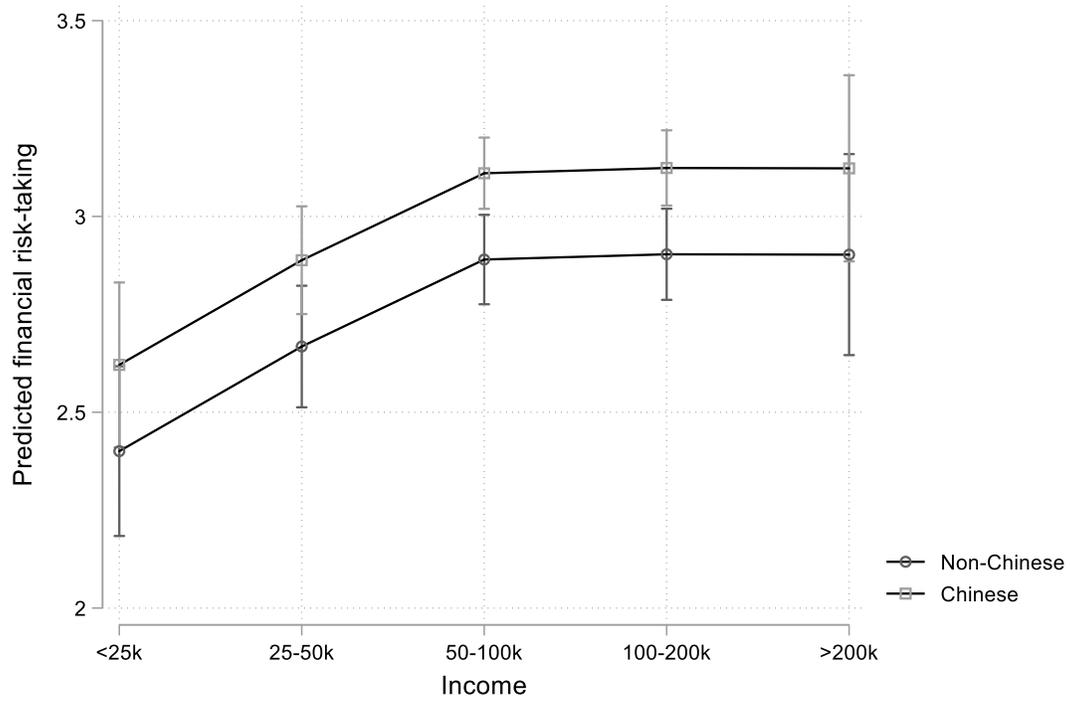

Figure 1. Risk-taking (DOSPERT-F scale) by ethnicity and income

Note: Predicted values based on an OLS regression of the DOSPERT-F scale on the covariates in Table 1, with covariates at observed values (regression estimates not reported). Capped bars indicate 90-percent confidence intervals.

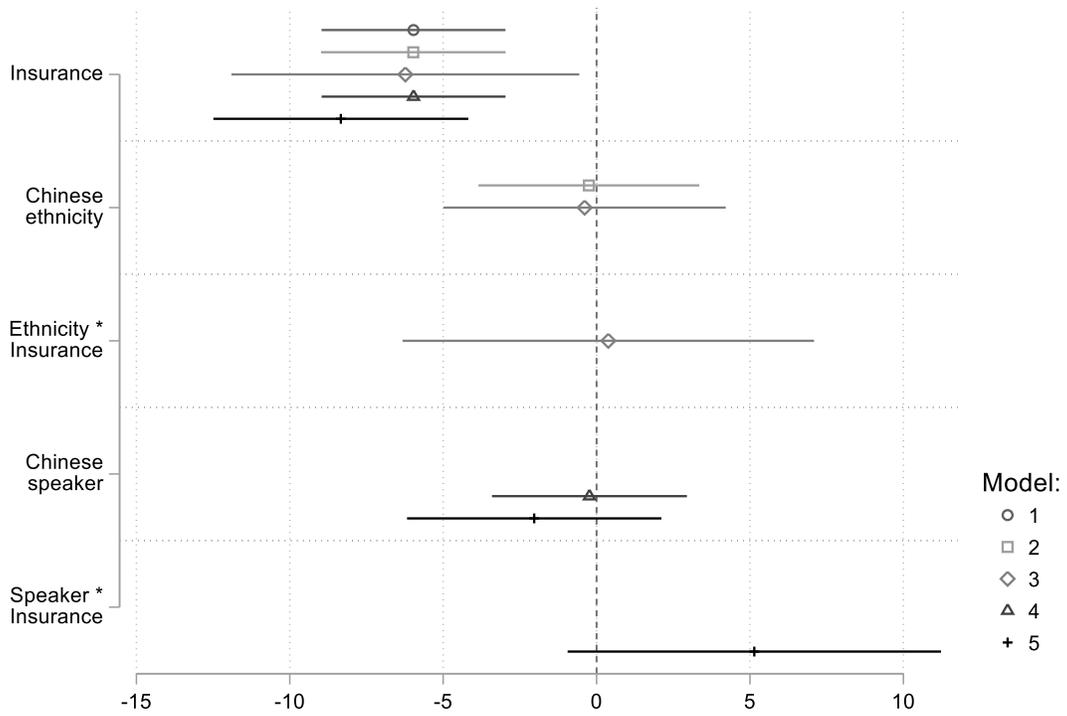

Figure 2. Insurance, ethnicity & language effects on savings

Note: OLS regression estimates (see Table 1). Horizontal lines indicate 95-percent confidence intervals.

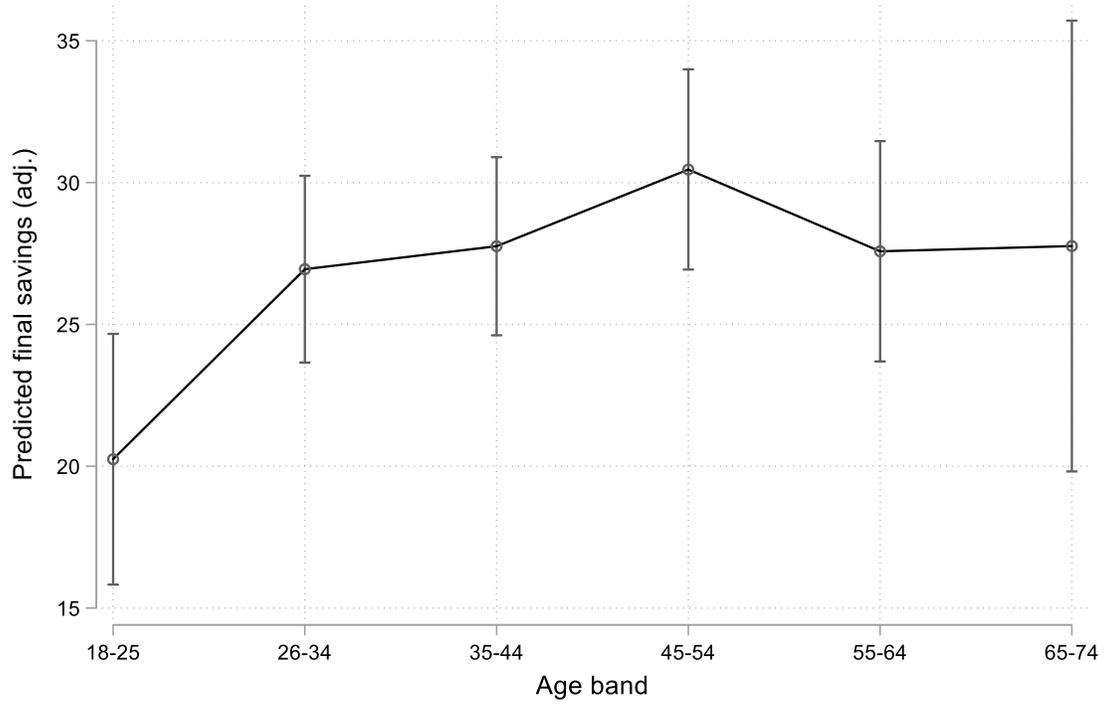

Figure 3. Savings by age

Note: Predicted values based on OLS regression estimates (see Table 1) with covariates at observed values. Capped bars indicate 95-percent confidence intervals. Predicted value for respondents 75 years and over omitted.

# Appendix

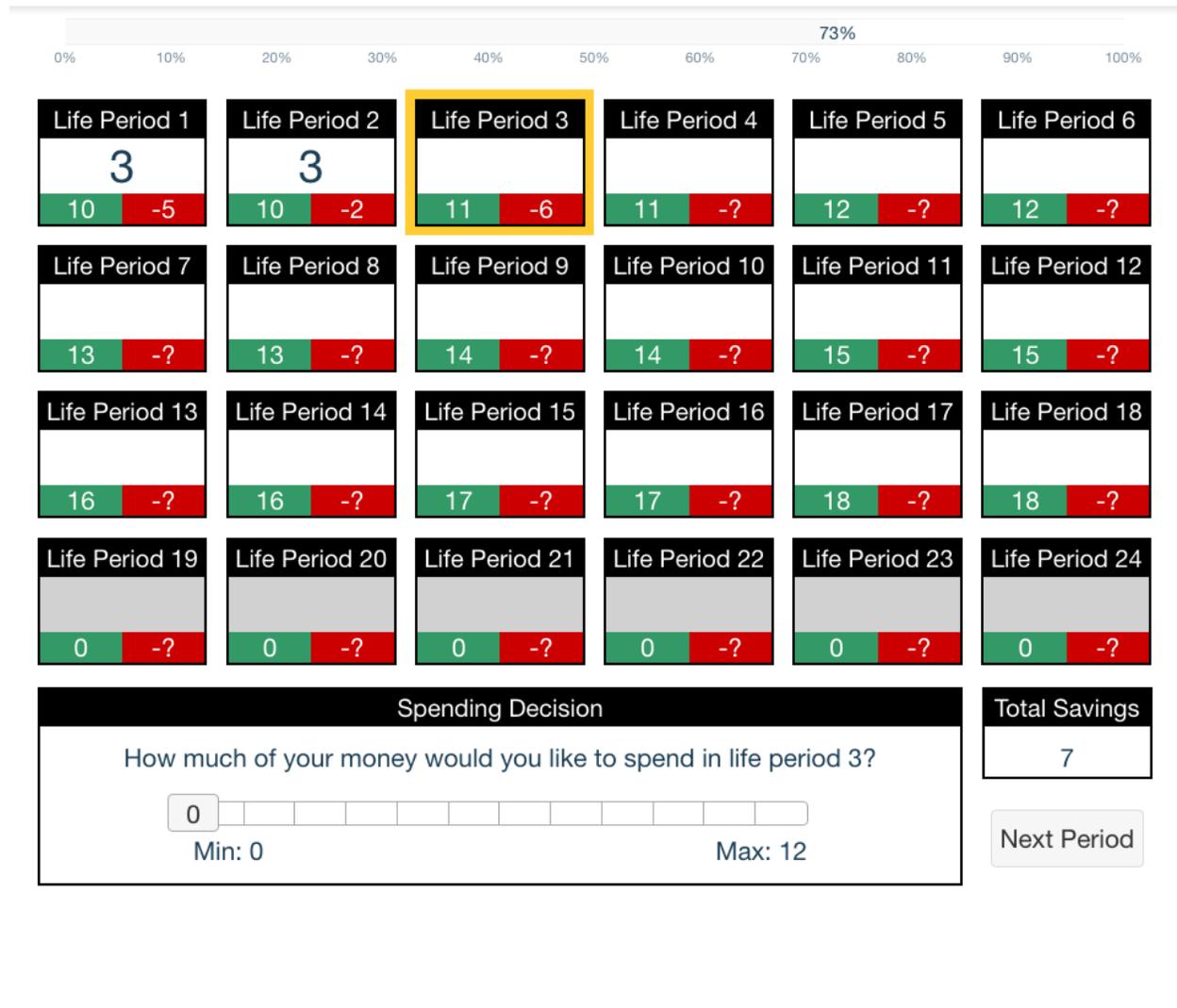

Screenshot showing an example round of the saving task in progress. The current period is highlighted in yellow. Income in each period is shown on a green background at the bottom left of each period box, while the expenses are shown in red at the bottom right of each box. Discretionary consumption in past periods is shown in black. Periods in grey are "retirement periods" without income. Total savings are shown at the bottom right.